\documentclass[twocolumn,secnumarabic,amssymb,amsfonts,nobibnotes,aps,prl,showpacs]
{revtex4}

\usepackage{amsmath}
\usepackage{latexsym}
\usepackage {mathrsfs}
\usepackage{float}
\usepackage{amssymb}
\usepackage{amsfonts}
\usepackage{graphicx}
\usepackage{textcomp}
\usepackage{hyperref}
\usepackage {mathrsfs}
 1

\newcommand{\dummy}

\begin{document}
\title{Transport Phenomena in Fluids: Finite-size scaling for critical behavior}
\author{Sutapa Roy and Subir K. Das$^{*}$}
\affiliation{Theoretical Sciences Unit, Jawaharlal Nehru Centre for Advanced Scientific 
Research, Jakkur P.O, Bangalore 560064, India}

\date{\today}

\begin{abstract} Results for transport properties, in conjunction with phase behavior 
and thermodynamics, are presented at the criticality of a binary Lennard-Jones fluid 
from Monte Carlo and molecular dynamics simulations. Evidence for much stronger 
finite-size effects in dynamics compared to statics has been demonstrated. Results for 
bulk viscosity are the first in the literature that quantifies critical divergence via 
appropriate finite-size scaling analysis. Our results are in accordance with the 
predictions of mode-coupling and dynamic renormalization group theoretical calculations. 
\end{abstract}

\pacs{64.60.Ht, 64.70.Ja}
\maketitle
\hspace{0.2cm} Understanding the properties of fluids is important from both basic 
research as well as technological point of view. Particularly, fluid behavior in the 
vicinity of critical point poses many interesting questions of fundamental importance 
\cite{Anisimov,Onuki,Ferrell,Olchowy,Fisher,Kim,Zinn,Hohenberg}. Fluids with short-range 
interactions, exhibiting gas-liquid and liquid-liquid transitions, are expected to have 
the static critical exponents $\beta=0.325,~\gamma=1.239,~\nu=0.63,~\alpha=0.11$, 
respectively, for order-parameter, susceptibility ($\chi$), correlation length ($\xi$) 
and specific heat, thus belonging to the three-dimensional Ising universality class 
\cite{Zinn}. On the other hand, it is expected that model H \cite{Hohenberg} should 
define the dynamic universality class for both the transitions. In dynamics the 
quantities of interest are shear ($\eta$) and bulk ($\zeta$) viscosities, thermal 
diffusivity and its analog, the mutual diffusivity ($D_{AB}$) in a binary fluid, the 
critical singularities for which are given by \cite{Onuki,Ferrell,Olchowy}
\begin {eqnarray}\label{dab}
D_{AB} \sim \xi^{-x_D},~ \eta \sim \xi^{x_\eta},~ \zeta \sim \xi^{x_\zeta};~ 
\xi \sim \epsilon^{-\nu},  
\end{eqnarray}
$\epsilon (={|T-T_c|}/{T_c})$ being a measure of the temperature $(T)$ deviation from 
the critical value $(T_c)$. In Eq. (\ref{dab}), exponents obey the scaling relations 
\cite{Onuki}
\begin {eqnarray}\label{xd}
x_D=1+x_\eta,~ x_\zeta=z-{\frac \alpha \nu};~x_\eta=0.068,~z=3.068,
\end{eqnarray}
where $z$ characterizes the divergence of relaxation time $\tau$, at $T_c$, with system 
size $L$ as
\begin {eqnarray}\label{tau}
\tau \sim {\xi^z} \sim L^z.
\end{eqnarray}
\par
\hspace{0.2cm} In addition to other static and dynamic properties, particular focus of 
this work is to understand the critical behavior of bulk viscosity that describes the 
response of a fluid to a compression or expansion. Even though the study of bulk 
viscosity is thought to be important for compressible fluids, the theoretical prediction,
 as our simulation results will also reveal, for similar critical enhancement for both 
gas-liquid (compressible) and liquid-liquid (incompressible) transitions is certainly 
interesting. While there are experiments \cite{Gillis} probing the critical behavior of 
$\zeta$, simulations are rare \cite{das2,Meier,Salin,Dyer} despite this being very 
important in the description of the damping of longitudinal sound waves. Dyer et al. 
\cite{Dyer} in fact pointed out the difficulty of studying bulk viscosity and suggested 
the need of significant effort to understand dynamics of continuous model fluids. The 
only noteworthy study of bulk viscosity in the context of criticality, so far, is due
to Meier et al. \cite{Meier} for gas-liquid transition of a single component 
Lennard-Jones (LJ) fluid who, however, did not quantify the critical divergence. While 
their data suffered from large error close to $T_c$, their observation of strong 
enhancement of $\zeta$ far above $T_c$ $(\simeq 4.5T_c)$, as is also observed by us, 
due to extremely slowly decaying pressure fluctuations, is very interesting. In fact, 
to the best of our knowledge, ours is the first simulation study of bulk viscosity that 
quantifies its critical divergence. In addition to confirming the theoretical predictions
 for critical exponents \cite{Onuki,Ferrell,Olchowy}, this work also provides direct 
evidence for stronger size effect in dynamics compared to statics. 
\par
\hspace{0.2cm} Apart from understanding the universality, computer simulations have been 
instrumental in providing many other details as far as static properties are concerned. 
In contrast, simulations of critical dynamics are very rare \cite{Jagannathan,S.K. Das,
Meier} the primary reason for which being the critical slowing down, as embodied in 
Eq. (\ref{tau}), that brings in additional complexity for the computation of dynamics 
over statics where finite-size effects are the only difficulty. Also, for molecular 
dynamics (MD) simulations in micro canonical ensemble, which is needed for perfect
preservation of hydrodynamics, it is extremely difficult to control the temperature to 
the desired value for a prolonged period of time due to truncation error. This may be a 
necessity at temperatures close to $T_c$ because of the presence of long-time tails 
\cite{Onuki,Alder}, particularly for quantities showing strong enhancement. All these 
problems combined together is suggestive of avoiding brute force method of simulating 
larger systems close to the critical point. The finite-size scaling method employed here 
to understand the results will demonstrate that if appropriate strategy is devised, all 
these hurdles could easily be overcome. Apart from critical phenomena, such methods 
could be useful in the study of other slow dynamics, e.g., glassy dynamics, where tools 
from critical phenomena are being recently adopted to understand growing dynamic length 
in supercooled liquids \cite{Biroli}.
\par
\hspace{0.2cm} We use a binary fluid ($A+B$) model \cite{S.K. Das,Allen} where, for 
$r_{ij}(|{\vec r_i}-{\vec r_j}|)< r_c$, particles at ${\vec r}_i$ and  ${\vec r}_j$ 
interact via \cite{S.K.Das,Allen} $u(r_{ij})=U(r_{ij})-U(r_c)-(r_{ij}-r_c)\left(dU/
dr_{ij}\right)_{{r_{ij}}=r_c},$ while $u(r_{ij}\ge r_c)=0$, with 
$U(r_{ij})=4\varepsilon_{\alpha\beta}[({\sigma}/{r_{ij}})^{12}
-({\sigma}/{r_{ij}})^6]$ $(\alpha,\beta \in A,B)$ being the standard Lennard-Jones (LJ) 
potential. For the choice $\varepsilon_{AA}=\varepsilon_{BB}=2\varepsilon_{AB}=
\varepsilon$,  we have a fully symmetric model that gives a demixing transition at 
\cite{S.K. Das} $T_c^{*}={k_BT_c}/{\varepsilon}=1.4230 \pm 0.0005$, for $r_c=2.5\sigma$. 
Phase diagram of such a system can be obtained from a semi-grandcanonical Monte Carlo 
(SGMC) simulation \cite{Landau}, where in addition to standard displacement trials, 
one introduces identity switch ($A \rightarrow B \rightarrow A$) moves, thus allowing 
for fluctuations in concentration $x_\alpha(={N_\alpha}/{\sum_{\beta}{N_\beta}})$ of 
species $\alpha$. The distribution $P(x_\alpha)$ of concentration fluctuation has 
double and single peak structures respectively at temperatures below and above $T_c$. 
While from the location of the peaks below $T_c$ one can obtain the phase diagram in 
${x_A}-T$ plane, static concentration susceptibility $(\chi)$ above $T_c$ can be 
calculated as ${k_B}T\chi={\chi^*}{T^*}=N({\langle {x_\alpha}^2 \rangle}-1/4)$, where 
the term $1/4$ corresponds to a critical concentration $x_A^c=1/2$ dictated by the 
symmetry of the model.
\par
\hspace{0.2cm} Transport properties were studied via MD simulations in microcanonical 
ensemble with $\zeta$ and the Onsager coefficient $\mathscr L(=\chi D_{AB})$ being 
calculated from Green-Kubo (GK) relations \cite{Allen,Hansen}
\begin{eqnarray}\label{zeta}
\zeta+{{\frac 4 3}\eta}={\left(\frac {t_0^3}{\sigma V{T^{^*}}m^2}\right)
\int_0^\infty dt 
\langle {\sigma^{'}_{xx}}(0){\sigma^{'}_{xx}}(t)\rangle},
\end{eqnarray}
\begin{eqnarray}\label{ons}
\mathscr L=\left(\frac {t_0}{N{T^{^*}}\sigma^2}\right) \int_0^\infty dt 
\langle {J_x^{AB}}(0){J_x^{AB}}(t) \rangle,
\end{eqnarray}
where $\sigma^{'}_{xx}=\sigma_{xx}
-P$, $\sigma_{xx}(=\sum_{i=1}^N[{m_i}{v_{ix}}{v_{ix}}+{\frac {1}{2}} \sum_j^{'}
{(x_i-x_j)F_{xj}}])$ being the diagonal elements of the stress tensor with 
$P=\langle{\sigma_{xx}}\rangle$ (GK formula for $\eta$ in Eq. (\ref {zeta}) contains 
the off-diagonal elements of stress tensor) and 
${J_x^{AB}}(t)(={{x_B}{\sum}_{i=1}^{N_A}{\vec v_{i,A}}(t)}
-{{x_A}{\sum}_{i=1}^{N_B}{\vec v_{i,B}}(t)})$ is the concentration current with 
$\vec {v_{i,\alpha}}(t)$ being the velocity of particle $i$ of species 
$\alpha$ at time t. In Eqs. (\ref{zeta}) and (\ref{ons}), $V$ is the volume of simulation
 box, $m$ is the mass of a particle and $t_0 [=(m\sigma^2/\varepsilon)^{1/2}]$ is the 
LJ time unit, which we set to unity. All results for dynamics are obtained from MD 
runs, with integration time step ${\Delta}t=0.005$, in a periodic cubic box of length 
$L$, in units of $\sigma$, after averaging over $160$ independent initial 
configurations.
\par
\begin{figure}[htb]
\centering
\includegraphics*[width=0.44\textwidth]{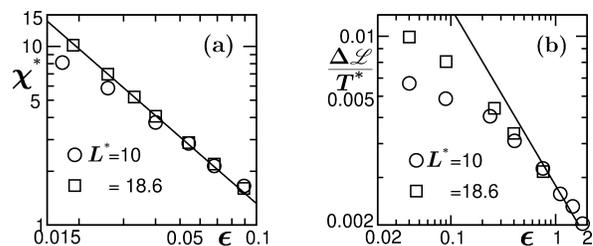}
\caption{\label{fig1} (a) Log-log plot of $\chi$ vs. $\epsilon$, for $L^*=10$, and 
$18.6$. The solid line represents the critical divergence of $\chi$ with exponent 
$\gamma=1.239$. (b) Plot of Onsager coefficient ${\Delta{\mathscr L}}/{T^*}$ vs. 
$\epsilon$ for $L^*=10,$ and $18.6$, on a log scale. Ths solid line here has critical 
exponent ${\nu_\lambda}=0.567$ and amplitude $Q=0.0028$.}
\end{figure}

\hspace{0.2cm} In Fig. \ref{fig1}(a) we show the results for $\chi$ as a function of 
$\epsilon$ for $L^*={L/\sigma}=10$ and $18.6$. The continuous line there has a power-law 
form with exponent $\gamma$ being fixed to its Ising value $1.239$. While this confirms
the Ising-like behavior, the consistency of the data for $L^*=18.6$ with the
solid line over the whole region is suggestive that finite-size effects did not appear 
yet. In view of the fact that finite-size effects were pointed out to be stronger in 
dynamics and a very strong background contribution was found in the study of mutual 
diffusion  \cite{S.K. Das}, we revisit it in the following.
\par
\hspace{0.2cm} In Fig. \ref{fig1}(b) we study the critical enhancement ${\Delta
{\mathscr L (T)}}[={\mathscr L}-{{\mathscr L}_b}]$ of Onsager coefficient, with 
${\mathscr L}_b$ being the contribution coming from short-range fluctuations and needs 
to be taken care of far above $T_c$, where critical enhancement is small. In the 
following we treat ${\mathscr L}_b$ as a constant, albeit weak temperature dependence 
that it might have. Here we plot $\Delta{\mathscr L (T)}$ which has the expected critical
 divergence 
\begin{eqnarray}\label{deltaL}
{\Delta{\mathscr L}}=QT\epsilon^{-\nu_\lambda}; ~~ {\nu_\lambda}=0.567,
\end{eqnarray} 
as a function of $\epsilon$, by adopting the constant value of ${\mathscr L_b}=0.0033$ 
as obtained in an earlier study \cite{S.K. Das}. Upon imposing \cite{S.K. Das} 
$Q=0.0028$ and ${\nu_\lambda}=0.567$, good consistency of the solid line is obtained 
with the simulation data, for large $\epsilon$. This, in addition to directly confirming 
the theoretical predictions as well as the conclusion drawn from the previous finite-size
 scaling study \cite{S.K. Das}, with very limited data, regarding the exponent 
and amplitude, is also indicative of a rather wide critical range. On the other hand, 
it is interesting to note from the comparison between Fig. \ref{fig1}(a) and Fig. 
\ref{fig1}(b) that size effects are appearing much earlier in $\mathscr L$ than in 
$\chi$, which requires appropriate attention to understand.
\begin{figure}[htb]
\centering
\includegraphics*[width=0.44\textwidth]{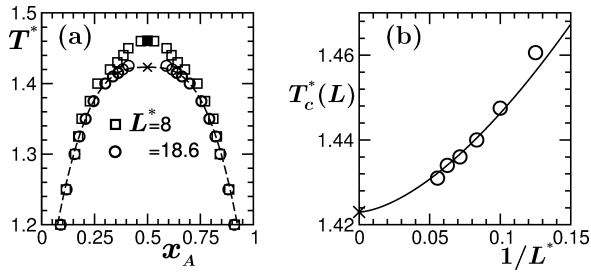}
\caption{\label{fig2} (a) Phase diagram of the model in ${x_A}-T$ plane for two system 
sizes. The filled symbol corresponds to $T_c^L$ for ${L^*}=8$, while the cross 
$(\times)$ locates ${T_c^{\infty}} \equiv {T_c}$. The dashed line there is a fit to the 
form $m =|{x_A}-1/2| \sim {\epsilon^\beta}$, taking data close to the critical 
point and unaffected by finite size. (b) Plot of $T_c^L$ vs. $1/L^*$ where 
the continuous line is a fit to the form (\ref{tcl}) with $\nu=0.63$, including only 
the four largest system sizes.} 
\end{figure}
With the knowledge about the spread of the critical region, we move forward to devise 
a strategy for a finite-size scaling analysis \cite{Kim,Landau,Fisherbook}, that will 
require only small systems and large temperatures so that difficulty due to long-time 
tails could be avoided. This will be tested with the better understood quantities, 
$\chi$ and $\mathscr L$ first, before applying it to $\zeta$.
\par
\hspace{0.2cm} As a first step, in Fig. \ref{fig2}(a) we show the phase behavior of the 
present model for different values of $L$ that exhibit strong size effect close to the 
critical point. We define a finite-size critical point \cite{Mon}, $T_c^L$, as the 
temperature where $P(x_A)$ in the SGMC simulation gets a single-peak structure from a
double-peak one with the increase of temperature. This is represented by filled symbol 
for $L^*=8$. Note that true meaning of a critical temperature can be assigned only when 
$L \rightarrow \infty$, which for the present case is marked by a cross and was obtained 
\cite{S.K. Das} in an unbiased manner from the method of intersection of Binder parameter \cite{Binder}. In Fig. \ref{fig2}(b) we demonstrate the variation of $T_c^L$ with $L$. 
The continuous line there is a fit (including data only for four largest values of $L^*$) to the expected scaling form in the large $L$ limit,
\begin{eqnarray}\label{tcl}
(T-{T_c^L}) \sim L^{-1/\nu}.
\end{eqnarray}
The deviation of data for $L^*=10$ and $8$ are due to corrections to scaling for small 
values of $L$, which can be numerically accounted for \cite{Fisher,Kim} by replacing 
$L^*$ by ${L^*}-\ell^*$ in the abscissa. Indeed for $\ell^*=2$ we get a perfect fit 
passing through all data points, which, in fact, has been used to obtain $T_c^L$ for 
intermediate $L$ values. Nevertheless, even for $L^*=8$, correction is very small and 
the simulation data deviates from the solid line by less than $1\%$ which is negligible 
compared to the thermal fluctuations during the MD runs.
\par
\begin{figure}[htb]
\centering
\includegraphics*[width=0.24\textwidth]{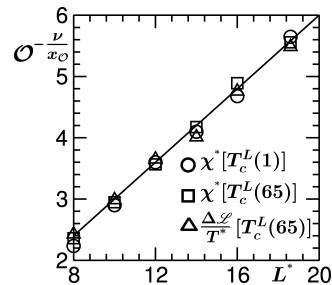}
\caption{\label{fig3} Plots of ${\chi^*}^{\nu/\gamma}$ and 
$(\frac{{\Delta}{\mathscr L}}{T^*})^{{\nu}/{{\nu}_{\lambda}}}$ vs. $L^*$ along different 
$f$-loci. Data for different quantities and $f$ values have been appropriately scaled 
to collapse. The continuous straight line is a guide to the eyes. }
\end{figure}

\hspace{0.2cm}At this stage, we define an effective finite-size critical point \cite{Kim} from ${T_c^L}[={T_c^L}(1)]$ as
\begin{eqnarray}\label {eff}
T_c^{L}(f)={T_c}+f({T_c^L}-T_c),
\end{eqnarray}
which has the same power-law convergence to $T_c$ as (\ref{tcl}). One can study critical 
behavior along different $f$-loci as a function of $L$, when, for an observable 
$\mathcal O$ $(\sim \epsilon^{x_{\mathcal O}})$, one obtains the scaling law
\begin{eqnarray}\label {eq16}
\mathcal O \sim L^{-{{x_{\mathcal O}}/{\nu}}},
\end{eqnarray}
where the amplitude will depend upon the value of $f$. Fig. \ref{fig3} demonstrates 
this for $\chi$ and ${\Delta{\mathscr L}}$ for two values of $f$, where we have 
plotted $\chi^{\nu/\gamma}$ and $(\frac{{\Delta}{\mathscr L}}{T})^{{\nu}/{{\nu}_
{\lambda}}}$ vs $L$. A linear behavior upon imposing $\nu=0.63$, $\gamma=1.239$ and 
$\nu_\lambda=0.567$  validates this strategy. Note that largest value of $f$ considered 
here is $65$ which gives ${T_c^{L}(65)}=3.875$ for $L^*=8$. Nice consistency of the 
data for whole range of $L$ is suggestive of only weak corrections even for the 
smallest $L$ considered.
\par
\begin{figure}[htb]
\centering
\includegraphics*[width=0.43\textwidth]{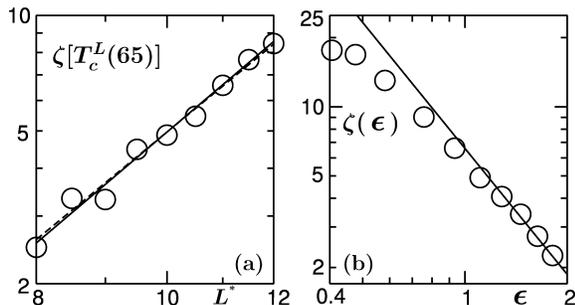}
\caption{\label{fig4} (a) Log-log plot of $\zeta$ as a function of $L$ at 
${T_c^{L}}(f)$ with $f=65$. The continuous line is a fit to $\sim L^{x_\zeta}$ giving
 $x_\zeta =2.96$ while the dashed line corresponds to an exponent $2.89$. (b) Plot of 
$\zeta$ vs $\epsilon$ for $L^*=10$. The solid line is a fit to the form
${A_\zeta}\epsilon^{-1.82}$ giving ${A_\zeta}=6.6\pm 1.0$.} 
\end{figure}

\hspace{0.2cm} Having demonstrated the usefulness of such method, encapsulated in Eqs. 
(\ref{eff}) and (\ref{eq16}), we adopt it to quantify the critical divergence of $\zeta$.
 Due to the technical difficulties to calculate it at lower temperatures for larger 
systems, in Fig. \ref{fig4}(a) we present it only for $f=65$. Very linear look of the 
whole data set on a log-log plot is suggestive of only small background contribution. 
Significant increase of $\zeta$ over only small range of ${L^*} \in [8,12]$, signals a 
strong divergence. The continuous line there is a fit to a power-law form 
$\sim L^{x_\zeta}$  giving ${x_\zeta}\simeq2.96$ which differs only by $2\%$ from the 
theoretical prediction (almost indistinguishable dashed line) 
${z-{\frac {\alpha}{\nu}}}\simeq 2.89$. Note that more recently \cite{Ferrell} it has 
been pointed out that the exponent is closer to $z$. While this confirms the expected 
theoretical behavior, we estimate the non-universal critical amplitude from the following
 exercise which will also provide a more direct confirmation of the exponent. In Fig.
 \ref{fig4}(b), we plot $\zeta$ as a function of $\epsilon$. Here, from our experience 
with $\mathscr L$, we choose a range with $\epsilon > 1$(that includes the last four 
points) for a fitting to the form $\zeta \sim \epsilon^{-\nu{x_\zeta}}$ by fixing 
$x_{\zeta}$ to $2.89$, which gives a critical amplitude $A_\zeta=6.6\pm 1.0$. Here the 
point of deviation of the simulation data from the solid line is consistent with the 
appearance of finite-size effect in $\mathscr L$. While the solid line provides an 
excellent fit to the selected region, an effective, though much smaller and misleading, 
exponent could also have been obtained from a fitting to the whole data set which has an 
average linear look on log-scale.
\par
\hspace{0.2cm} In summary, dynamic critical phenomena is studied in a symmetric binary 
fluid. Consistency with predictions of dynamic renormalization group and mode-coupling 
theories has been established. Quantitative understanding of the bulk viscosity via 
computer simulation is the first in the literature. Critical region appears to be rather 
wide so that with appropriate application of finite-size scaling method it has been 
possible to stick to only small systems at large temperatures. A possible reason for 
stronger finite-size effects in dynamics compared to statics could be back-flow due to 
periodic boundary conditions \textminus~however, significant attention is required to 
settle this important issue.
\par
\textit{Acknowledgment}: SKD acknowledges previous fruitful collaboration with  M.E.
 Fisher, K. Binder, J.V. Sengers and J. Horbach. He also thanks K. Binder, J.V. Sengers 
and J.K. Bhattacharjee for critical reading of the manuscript and useful comments. The 
authors acknowledge grant number SR/S2/RJN-$13/2009$ of the Department of Science and 
Technology, India. SR is also grateful to CSIR, India, for financial support.
\par
\hspace{0.2cm}${*}$ das@jncasr.ac.in

\end{document}